\documentclass[letterpaper]{article}
\usepackage{pdfpages}
\usepackage{dcolumn}
\usepackage{scrlayer-scrpage}
\usepackage{graphicx}
\usepackage{aaai}
\usepackage{times}
\usepackage{helvet}
\usepackage{courier}
\usepackage[T1]{fontenc}
\usepackage[english]{babel}
\title{Caveat Emptor, Computational Social Science: Large-Scale Missing Data in a Widely-Published Reddit Corpus}
 \author{Devin Gaffney \\ Network Science Institute, Northeastern University \\ 177 Huntington Avenue \\ Boston, MA, 02115
 \And
 J. Nathan Matias \\ Princeton University \\ Peretsman-Scully Hall \\ Princeton, NJ, 08540}

\begin{document}

\maketitle

\section{Abstract}
As researchers use computational methods to study complex social behaviors at scale, the validity of this computational social science depends on the integrity of the data. On July 2, 2015, Jason Baumgartner published a dataset advertised to include ``every publicly available Reddit comment'' which was quickly shared on Bittorrent and the Internet Archive. This data quickly became the basis of many academic papers on topics including machine learning, social behavior, politics, breaking news, and hate speech. We have discovered substantial gaps and limitations in this dataset which may contribute to bias in the findings of that research. In this paper, we document the dataset, substantial missing observations in the dataset, and the risks to research validity from those gaps. In summary, we identify strong risks to research that considers user histories or network analysis, moderate risks to research that compares counts of participation, and lesser risk to machine learning research that avoids making representative claims about behavior and participation on Reddit.

\section{The Baumgartner Reddit Corpus}
Trace data sourced from online platforms has become an essential component for many forms of research ranging from sentiment analysis \cite{pak2010twitter} to epidemiological modeling \cite{abdullah2011epidemic} and economics \cite{bollen2011twitter}. Dominant social platforms such as Twitter and Facebook have provided researchers with opportunities to directly study complex phenomena that, at their root, rely strongly on the nature of social interaction \cite{bond201261}. The reason for this, as \citeauthor{tufekci2014big} argues, is that large platforms (specifically Twitter, in this analogy) serve as a \textit{model organisms} for the social sciences, ones that allow for ideal conditions for measurement of many phenomena in a relatively accessible form.

On July 2, 2015, a new model organism was provided to researchers by Jason Baumgartner -- a ``complete'' copy of one of the largest forums, Reddit, which has gained high visibility in the past several years due to events such as the Reddit blackout \cite{matias2016going,newell2016user,baumgartner2015dataset} and the Gamergate controversy \cite{massanari2015gamergate}. Subsequently, many researchers have adopted the dataset, and have used it to study a wide range of questions, including the evolution of social networks \cite{fire2016analyzing}, user migration through online platforms \cite{tan2015all,newell2016user}, hate speech \cite{saleem2016web}, and online behavior research methodology \cite{barbosa2016averaging}, among others. 

As a social news platform, Reddit hosts discussions about text posts and web links across hundreds of communities called ``subreddits'' \cite{leavitt_upvoting_2014,massanari_participatory_2015} Discussions from public subreddits are aggregated by a variety of news aggregators to create the ``front page of the web'' that Reddit was founded to provide to its readers \cite{massanari__2015}. While the site also provides chatrooms and features for live discussions of breaking news \cite{leavitt_role_2017}, the most common reddit experience is centered around top-level \textit{submissions} and the \textit{comnents} that people post when discussing those submissions within their subreddit communities. The Baumgartner dataset follows this common experience and includes submissions and comments.

Researchers are drawn to the Baumgartner Reddit dataset for its completeness. In principle, a complete dataset improves research validity by avoiding the ambiguities of samples provided by platform application programming interfaces (APIs) and third-party data resellers \cite{lotan2011arab,diaz_online_2016}. In this paper, we show that this dataset, as distributed and used by researchers, is not as complete as reported. We report on gaps in this data, categorize the risks to research validity from these gaps, and share collaborative re-analyses of peer-reviewed papers that have used this dataset. Finally, we conclude with reflections on the sensitivity of online behavioral research to the kinds of gaps we found in the Baumgartner Reddit Dataset.

\subsection{Sequential ID Analysis}

The Baumgartner Reddit dataset came about through a convergence of factors: a mostly-public conversation platform, engineering details specific to the design of the Reddit system, and a creative data scientist who capitalized on these characteristics to contribute a unique dataset to public knowledge.

Many databases include the concept of an Identity column, or a column that generates an internal ID to serve as a unique reference to the row, or object, within the database. In many cases, this value auto-increments -- the first value in the database assumes a value of 1, the next, a value of 2, and so forth. This number can be artificially shifted within the space -- for instance engineers may partition early IDs of 1-1,000,000 for experimenting with data, for some reason, and start all production-system data created by users with ID 1,000,001. Aside from this possibility, if an object contains an ID of \textit{n}, then it is plausible to assume that there are at least \textit{n} objects within the database. 

In personal correspondence, Baumgartner  explained that this intuition led him to develop systems designed to systematically-collect all data on Reddit. Baumgartner's algorithm batches up 100 integers, converts them to the Base 36 representation that Reddit uses to represent their objects, and then queries for those objects. Reddit then returns the request with a set of all public, found objects. Baumgartner's algorithm can be run in a highly parallel environment -- many batches of 100 IDs can be concurrently requested, with no need to interact with one another. On other platforms, some error may be returned if data has been deleted. With Reddit, no error is returned -- instead, a truncated object reflecting that this deletion has occurred is returned. Therefore, barring technical issues, this method should provide a complete accounting for every ID within the range 1-\textit{n} for all public comments and submissions within the dataset. Using this method, Baumgartner archived the public record of reddit comments and submissions from the platform's creation through July 2015. Baumgartner has continued to provide this data as a freely-available resource.

In this analysis, we consider The full dataset as released by Baumgartner in July 2015, supplemented with updates published by Baumgartner through the end of February 2016. We also include a followup analysis extended to June 2017.


\section{Diagnosing Missing Data}
Because reddit comments and submissions have unique, sequential IDs, we can analyze gaps in the sequence to evaluate the completeness of the dataset. We observed two kinds of missing information: dangling references (known unknowns) and gaps indicated by the absence of information that we would expect to exist given the use of sequential integers to index comments and submissions (unknown unknowns).


We discovered the completeness problem when working with this dataset for our own research. Taking a random sample of subreddits and generating a timeline of daily comments and submissions, plots showed impossible results given the architecture of Reddit: some comment timelines started earlier than their corresponding submission timelines. 

The first kind of gap we discovered were dangling references. On Reddit, comments can only occur within a discussion of a submission and can only refer to other comments or submissions. In all cases, a submission would have to exist for a comment to refer to it, a relationship that is unidirectional in time. By traversing these relationships, we observed many references to missing comments and submissions. These can be thought of as ``known unknowns:'' comments which refer to other comments or to a parent submission, where the referred-to comment or parent submission is not contained within the Baumgartner dataset.

We also observed a second kind of gap: objects that are never referenced in the dataset but are likely missing. If all comments and submissions are given sequential integer IDs, we would expect an unbroken sequence of integers to be associated with information in the dataset. This is not the case. Consider the comments dataset: the earliest comment in the  Baumgartner dataset is comment \#2 and the highest is \#29,484,960,643. In October 2007, the Reddit Company incremented the comment IDs by several billion IDs. When accounting for this difference, we assume that any other gaps in the sequence of comment IDs can be attributed to gaps in the dataset: we count 943,755 total potentially-missing comments up to February 2016.

Missing comment IDs could be attributed to many possible causes. These IDs could be dangling references, public information that for some reason were not returned by Reddit's systems to Baumgartner's software at that moment, or information that was part of a community that had set its discussions to be private. These missing IDs are not associated with deleted content, since the Reddit platform returns information about deleted data, which is included in the Baumgartner dataset.

For submissions, we are less confident about the magnitude of missing unknown unknowns. While we have observed 1,539583 ``gaps'' in the space of IDs for submissions through February 2016, the first submission in the Baumgartner dataset starts at 9,970,002. When searching for submissions between \#1 and \#9,970,001, we have successfully found some submissions, leading us to believe that  millions of submissions from the early history of Reddit may be absent from this dataset.


\begin{table}
  \small
  \begin{center}
    \begin{tabular}{ |c|c|c| } 
      \hline
      Data Type & Comments & Submissions \\
      \hline
      Dangling References (to Feb 2016) & 101,257 & 405,911 \\
      Unknown Unknowns (to Feb 2016) & 943,755 & 1,539,583 \\
      Unknown Unknowns (to Jun 2017) & 35,801,325 & 27,795,423 \\
      \hline
    \end{tabular}
  \end{center}
  \caption{Totals for missing data in the Baumgartner dataset}
  \label{table:total_missing}
\end{table}

\begin{figure}[h]
  \centering
  \includegraphics[width=1\linewidth]{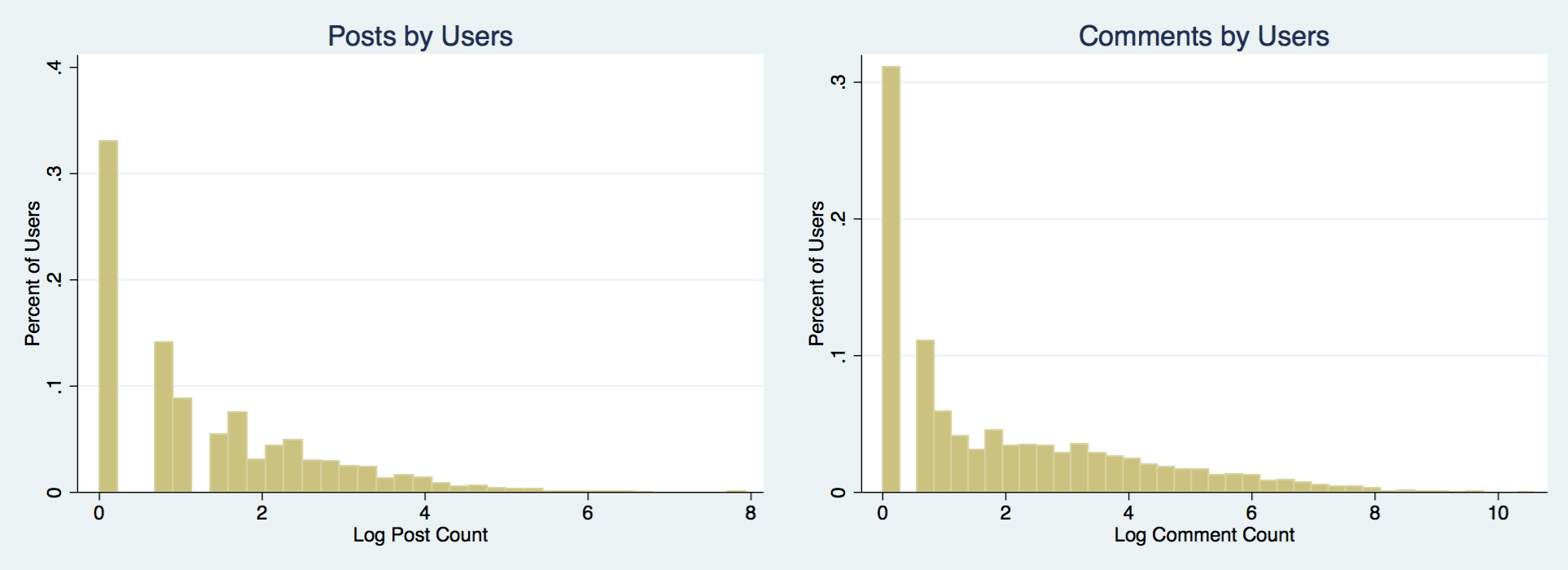}
  \caption{Histograms of sampled user Submission and Comment counts}
  \label{fig:post_counts}
\end{figure}

Deleted content, which is included in this dataset, represents a risk to validity that we do not consider here. A user who deletes even one comment in their posting history introduces many of the problems we describe in this paper, even if the fact of the comment is recorded in the Baumgartner dataset.


\subsection{The Per-User Risk of Missing Data}
How likely is a researcher to encounter these gaps? To address this question, we estimate the per-user risk of missing data, using a random sample of 7,400 accounts from the Baumgartner dataset. 

The average user in this sample commented 6.8 times and commented 96.6 times from late January 2006 through February 2016. These averages occur on a highly skewed distribution, as illustrated by the log-histograms in \ref{fig:post_counts}. Based on table \ref{table:total_missing}, the known maximum amount of missing comments and submissions is 943,755 and 1,539,583, respectively -- dangling references are a subset of ``unknown unknowns.'' Across the entire Baumgartner dataset, only 0.043$\%$ and 0.65$\%$ of comments and submissions, respectively, are missing. The issue has a compounding effect, since a small number of users create a large amount of the content on the platform. The more posts and comments someone produces, all else being equal, the more likely their histories will be affected by the missing data issue. As we have also showed, unknown unknowns expanded dramatically in the 16 months following February 2016 and now include 36 million missing comments and 28 million missing submissions.

What is the probability of data loss for an individual Redditor history? While in reality the missing data is not uniformally distributed throughout the corpus, we can estimate the effect by compounding probabilities to assess the degree to which a user could be affected by only a small amount of missing data. Using the averages from earlier, we can calculate the risk of any individual submission $r_s$ or comment $r_c$ being missing  simply by $\sum_c^n r_c$ and $\sum_s^n r_s$, respectively. In this case, the ``average'' Redditor may be exposed to a total maximum risk level of $\propto$ 4.18$\%$ likelihood for missing at least one comment and $\propto$ 4.46$\%$ for missing at least one submission. In the 7,400 individual set, approximately 2$\%$ of the sampled users had a 50$\%$ or greater chance of having a missing comment, and 2.6$\%$ of the sampled users had a 50$\%$ or greater chance of having a missing submission. These estimates were based on the census of dangling references and unknown unknowns from the beginning of the corpus to February 2016; we expect relatively similar rates in later data, since the rate of missed content has been consistent for the past several years. We offer these rough approximations to communicate a qualitative sense of how this missing data issue may create an appreciable problem for some forms of research. We include a more detailed typology of possible errors below.

\subsection{Distribution of Gaps Across Time}

\begin{figure}[h]
  \centering
  \includegraphics[width=1\linewidth]{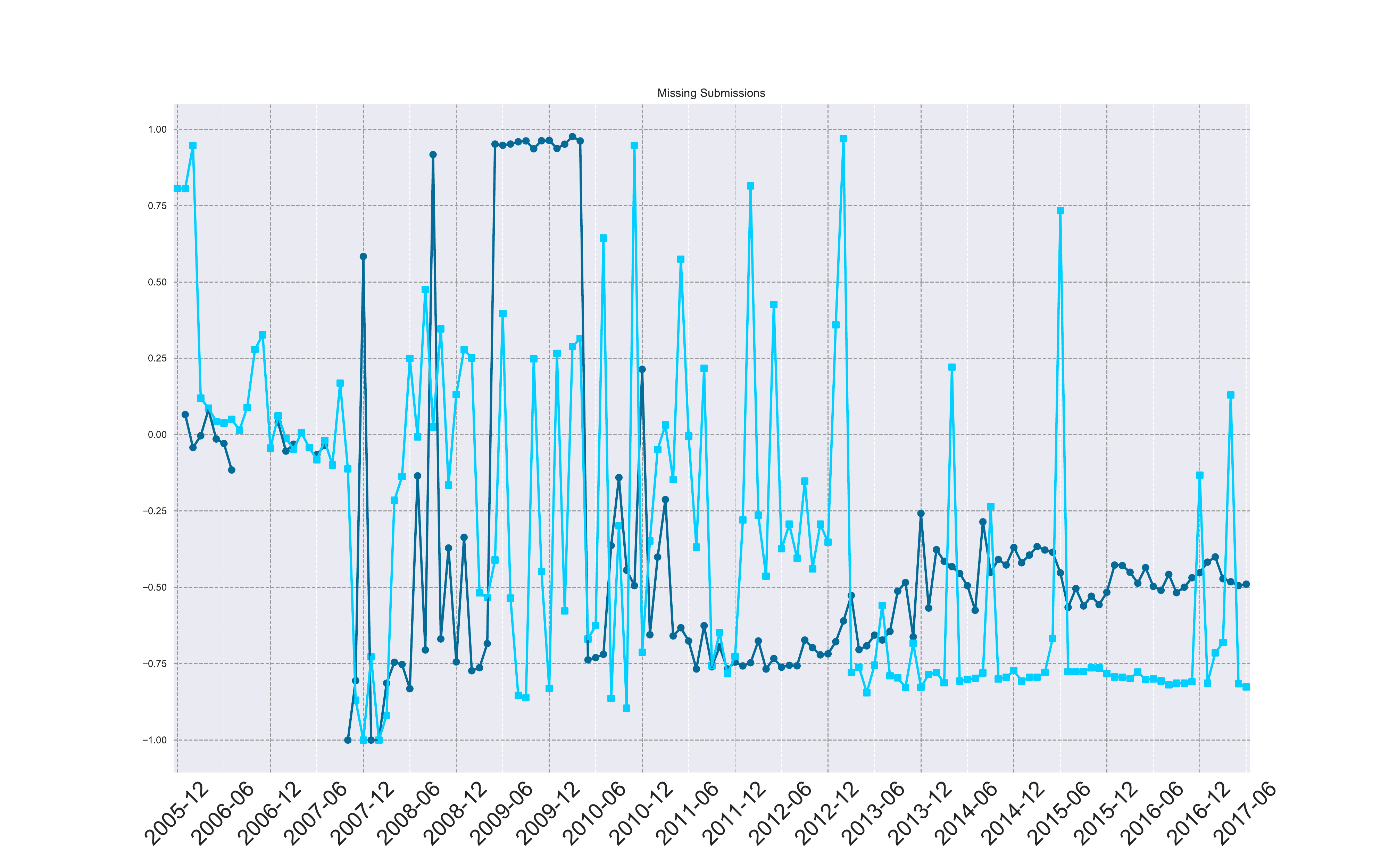}
  \caption{Burstiness of missing submissions and comments per month, 2005-June 2017}
  \label{fig:burstiness}
\end{figure}

Far from being uniformally distributed throughout the dataset, the instances of missing data appear to be ``bursty'' -- clustered at certain moments of time. Consequently, certain spaces in the Reddit network or certain time periods may be at greater risk of missing data than others. Importantly, we found significant gaps for comments at key moments in Reddit history that have been subjects of research, including the SOPA/PIPA protests \cite{benkler2015social} and the months leading up to the Reddit blackout \cite{matias2016going}. Leaning on \citeauthor{jo2012circadian}, we employ a measure of ``burstiness'' which considers the relative dispersion of errors throughout the ID space per each month of gathered data. This measure is bounded from [-1,1], where a score of -1 indicates completely evenly dispersed errors, and a score approaching 1 indicates that errors are located in a more concentrated set of missing blocks. Figure \ref{fig:burstiness} shows many high positive burstiness scores, indicating that missing blocks are often distributed unevenly within months throughout the dataset. 


\begin{figure}[h]
  \centering
  \includegraphics[width=1\linewidth]{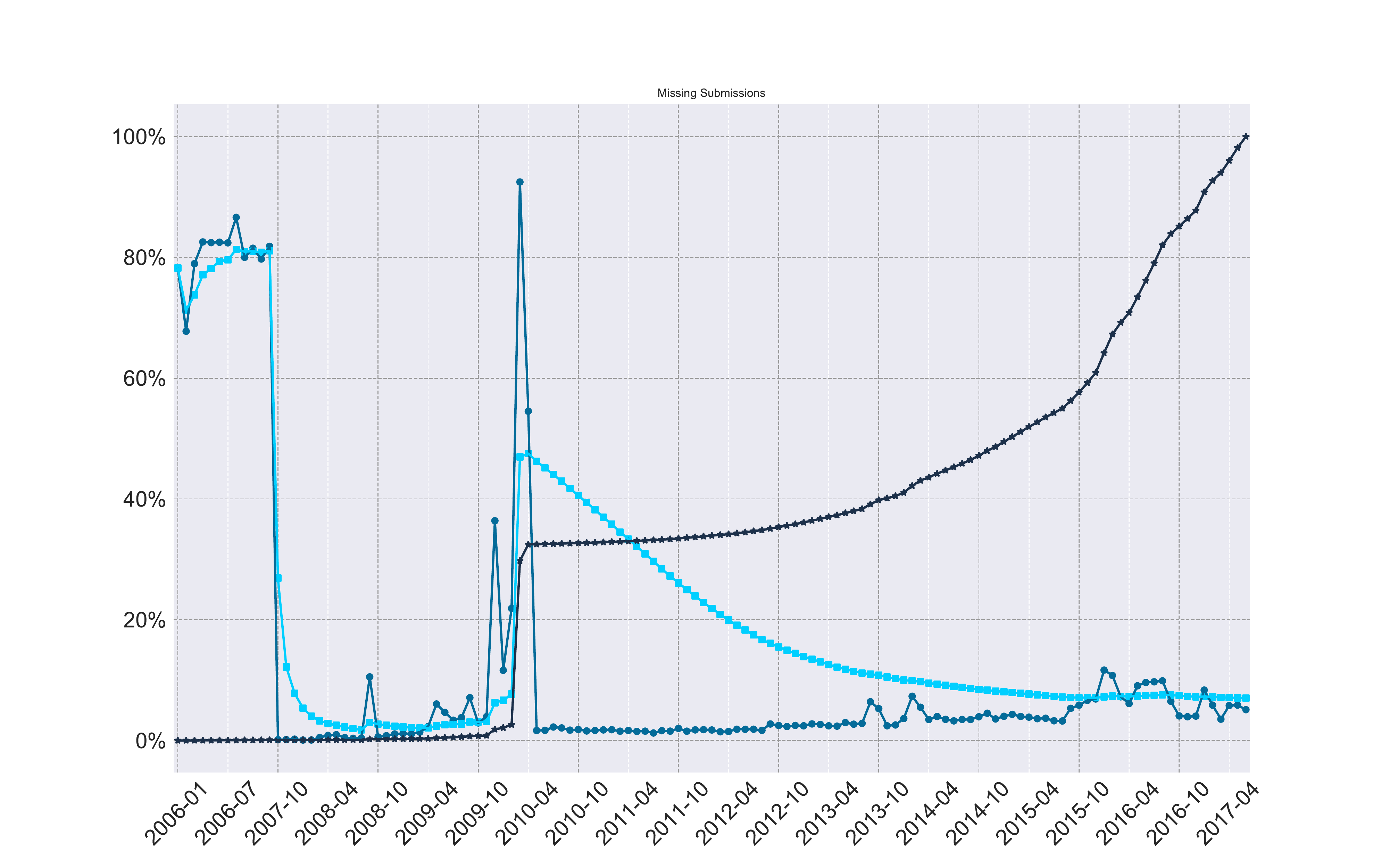}
  \caption{Varied measures of missing submissions per month. Medium blue circles denote the percent of submissions missing for each month of data, bright blue squares denote the rolling average percent of missing submissions, and dark blue stars denote the cumulative total number of missing submissions to date.}
  \label{fig:submissions_per_month}
\end{figure}

\begin{figure}[h]
  \centering
  \includegraphics[width=1\linewidth]{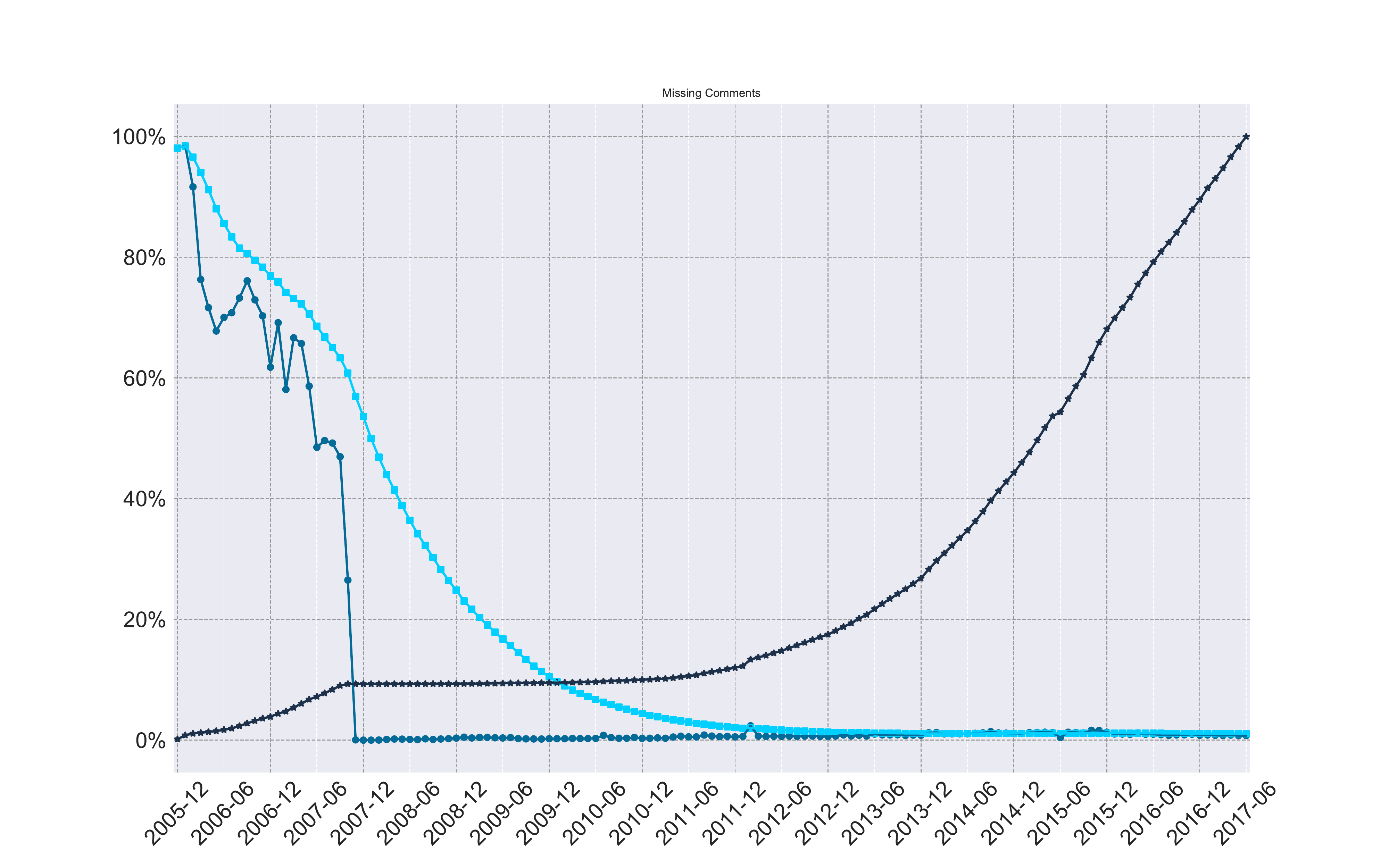}
  \caption{Varied measures of missing comments per month. Medium blue circles denote the percent of comments missing for each month of data, bright blue squares denote the rolling average percent of missing comments, and dark blue stars denote the cumulative total number of missing comments to date.}
  \label{fig:comments_per_month}
\end{figure}

\begin{figure}[h]
  \centering
  \includegraphics[width=1\linewidth]{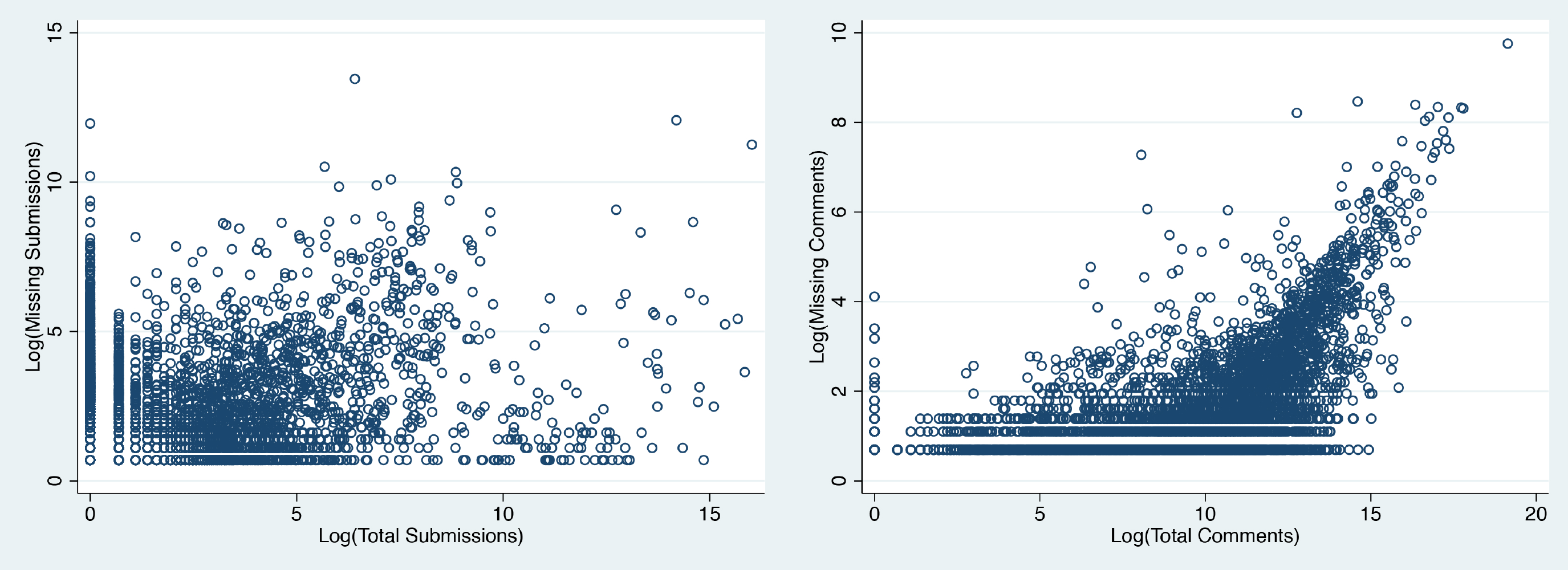}
  \caption{Gaps are not evenly distributed across communities. The total historical counts of comments per community comments are mildly correlated with the number of dangling references, while submissions are not very correlated with the number of dangling references.}
  \label{fig:scatter_full}
\end{figure}

Overall, figures \ref{fig:submissions_per_month} and \ref{fig:comments_per_month} illustrate an initially erratic distribution of errors throughout the dataset. These errors appear to occur directly within periods of substantial research interest and may affect several published results \cite{matias2016going,newell2016user}. While the rate of error was particularly erratic in early years, and the distribution of errors per ID gap continues to be erratic (figure \ref{fig:burstiness}), the error rate per month has evened out to around 1$\%$ missing data per month. 

\subsection{Distribution of Gaps Across Communities}

\begin{table}[!htbp] \centering 
\small
\begin{tabular}{@{\extracolsep{5pt}}lD{.}{.}{-3} D{.}{.}{-3} } 
\\[-1.8ex]\hline 
\hline \\[-1.8ex] 
\multicolumn{1}{c}{Variable} &
\multicolumn{1}{c}{Submissions} & \multicolumn{1}{c}{Comments}\\ 
\hline \\[-1.8ex] 
 Total Content & 0.212^{***} & 0.217^{***} \\ 
& (0.008) & (0.006) \\ 
  Month Subreddit & 0.005^{***} & 0.002^{**} \\ 
  \quad Created & (0.001) & (0.001) \\ 
  Constant & 1.198^{***} & -0.518^{***} \\ 
  & (0.094) & (0.095) \\ 
 \hline \\[-1.8ex] 
Observations & \multicolumn{1}{c}{8,176} & \multicolumn{1}{c}{4,341}  \\ 
R$^{2}$ & \multicolumn{1}{c}{0.086} & \multicolumn{1}{c}{0.306} \\ 
Adjusted R$^{2}$ & \multicolumn{1}{c}{0.086} & \multicolumn{1}{c}{0.305} \\ 
\hline 
\hline \\[-1.8ex] 
\textit{Note:}  & \multicolumn{2}{r}{$^{*}$p$<$0.1; $^{**}$p$<$0.05; $^{***}$p$<$0.01} \\ 
\end{tabular} 
  \caption{Regression exploring the relationship between amount of missing content per subreddit and total amount of known content per subreddit, and month in which the subreddit was created. We expect that these two variables would have meaningful explanatory power for where missing content is -- we find that this appears to be the case for missing comments but not for missing submissions, as evidenced by the relative $R^2$ values.}
  \label{fig:regression_results} 
\end{table} 

We also considered the degree to which missing content differentially affects individual subreddits. If data from some communities were more affected by gaps than others, the gaps could influence the results of comparative research about populations communities \cite{hill_studying_2017}.  If gaps affected communities equally, we would expect that the number of missing pieces of content monotonically rises with the number of overall pieces of content posted to a subreddit. As figure \ref{fig:scatter_full} shows, we find only marginal evidence for such a supposition. While more missing content is positively and significantly associated with larger subreddits, we do not find a direct relationship. One confounding factor may be the temporal ``center of gravity'' of a subreddit -- older subreddits are positioned at a time when more content was missing, on average, which may differentially affect older subreddits. We attempted to control for subreddit age in a multiple linear regression which accounted for the size of subreddits as well as the time at which those subreddits were created; we did not find any meaningful increase in explanatory power in the adjusted model. Table \ref{fig:regression_results} . The time at which a subreddit was created, however, is a poor proxy for the true ``center of gravity'' of content (i.e. the time at which a subreddit was most active), a characteristic that these models do not account for.

In the above sections, we have considered the influence of potentially-missing content on analyses of users, behavior over time, and groups. We observed numerous sources of potential bias in research: a substantial percentage of users could be affected by these gaps, the gaps are not evenly distributed across time, and gaps are not evenly distributed across communities. 

\section{How Missing Data Affects Common Research Methods in Computational Social Science}
How might these gaps influence research in practice? We expect that researchers asking different kinds of questions will face different kinds of risks from missing data. In the following sections, we categorize published literature that uses  this dataset and offer a typology of the risks that these gaps represent to common research methods in computational social science.

\textit{User history analysis} papers face the \textit{highest risks} from missing data, since a missing comment or submission could hide an important part of that user's history. A network analysis may fail to include a user's participation in a particular community or interaction with a key user. Furthermore, survival analyses might mis-estimate the moment of a person's departure or their participation level. \textit{Network analysis} papers also face \textit{high risks}, since the presence or absence of a tie could be dependent on the missing data. \textit{Sum analyses} that count the size or incidence rate of participation in subreddits or the use of certain kinds of language face \textit{moderate risk}, especially when analyzing small communities and rare events. \textit{Content analysis} that involves training machine learning systems on Reddit comments face \textit{minimal risk} because their research rarely includes claims about the population of Reddit users.

\subsection{Risk to User History Analyses}

Papers that test hypotheses based on user histories on Reddit may have substantial gaps in the histories that they seek to test. Analyses on user histories that consider the history in full are, in general, exposed to the highest risk -- analyses that are especially sensitive to high-volume users are very likely, on average, to consider users whose histories have gaps. \citeauthor{hessel2016science}, for example, observes and compares sums of comment participation between subreddits, and observes the full chain of user history -- \citeauthor{hessel2015democrats} adopts a similar approach. \citeauthor{barbosa2016averaging} compares year cohorts of individual-level behavior across all of Reddit, and as has been shown, some years are more affected by gaps than others. Additionally, the large number of potential missing submissions from Reddit's earliest years may also affect these findings. If a user history analysis requires the complete posting history between subreddits for a given user, gaps in such transmissions may constitute meaningful gaps in explaining a wide array of hypotheses.

\subsection{Risks to Network Analyses}

Some papers test network hypotheses by constructing interaction networks between users or communities, sometimes over time. Data gaps also represent a high risk to these papers, since missing submissions may result in unobserved ties in the network. \citeauthor{tan2015all} observes histories of user accounts participating in different communities, while \citeauthor{fire2016analyzing} observe network ties over time modeled on user histories. Substantial blocks of missing data, including the potentially large amount of missing submissions from Reddit's nascency could redraw the map of community ties on the platform. Tree structures reconstructing threads are also similarly affected, such as work by \citeauthor{hessel2017cats} and \citeauthor{fire2017rise}, which through linkages of comments and submissions similarly face issues due to missing submissions (i.e. parents of threads) or comments.

\subsection{Risks to Research That Counts and Compares Participation Between Communities}

Other papers test hypotheses based on participation sums within communities. Gaps that are biased toward particular communities will represent a risk to the validity of these studies. \citeauthor{matias2016going} observes levels of subreddit participation by moderators, observes relative participation levels of subreddit commenters in other subreddits, and observes moderator participation in ``metareddits''. \citeauthor{newell2016user} observes comment volumes within subreddits. \citeauthor{pew2016reddit} observes comments about political candidates across Reddit during a period where many submissions are within the dataset. \citeauthor{barbaresi2015collection} analyzes German language text to identify relative commenting rates about places in Germany. 
\citeauthor{horne2017impact} considers posts within /r/worldnews to determine linguistic characteristics of why some news frames are more visible than others. \citeauthor{dosono2017exploring} considers a specific set of communities associated with self-expression of Asian-American Pacific Islander (AAPI) identity on the platform.

As we showed in figure \ref{fig:scatter_full}, gaps do not appear to be evenly distributed across communities, since the number of missing comments and submissions per community is not strongly correlated to the number of observed comments and submissions in that community. While a simple statistical regression between the total counts of missing data and known data shows the relationship to be significant, the $R^2$ is low enough in both cases to lead us to conclude that studies on some subreddits could lead towards very biased results due to higher than random amounts of missing data. 

In practice, we observe 78 subreddits where at least 20$\%$ of the comments are missing, and 1,755 subreddits where at least 20$\%$ of the submissions are missing. Among subreddits that have any dangling references, on average they are missing at least 35$\%$ of their submissions. The $R^2$ score in a model predicting the volume of a community's missing observations from the volume of observed comments and submissions only explains 30$\%$ of the variance of missing comments and 10$\%$ of the variance of missing submissions (figure \ref{fig:scatter_full}). The risk to any specific study will depend on the distribution of gaps across the specific communities being compared.

\subsection{Risks to Machine Learning Models}

Finally, some studies train machine learning models and conduct linguistic analysis of the Baumgartner dataset. Insofar as these studies do not make claims about populations, gaps represent a minimal risk to the validity of this research. For example, \citeauthor{saleem2016web} trains machine learning models on comments from particular subreddits that have since been quarantined or banned by Reddit for harmful behavior. 

In our observations of communities where the mass of missing data is pooled, it seems to trend towards such communities -- across the three subreddits considered in their work, one of those subreddits has a large number of dangling references: observed comments refer to 696,642 unique missing submissions in the dataset for this one community alone. Among comments, 1,100 of 1,585,014 total comments were known to be missing. Saleem, Dillon, Benesch, and Ruths have re-analyzed their data after filling some gaps and fail to find any substantial differences in the performance of their machine learning models (citation forthcoming). Furthermore, since the purpose of this kind of machine learning research is to make inferences about out-of-sample observations rather than to test hypotheses about a population, such research may be less sensitive to variation due to missing data.


\section{Discussion}

All datasets have biases, no matter how complete we wish them to be. In the process of designing research, conscientious researchers will study those biases, document them, and account for them as best as possible. In this paper, we have shown ways in which an influential public dataset does not represent the ``complete'' record that its publisher and users aspired to. We have documented per-user risks of missing data, risks from the uneven distribution of missing data over time, and risks in the uneven distribution of missing data across communities. We have outlined the risks to research validity represented by these data gaps, including some of our own work.

We have raised these issues in direct conversation with Baumgartner, who has quickly and graciously re-processed ID blocks with missing data and filled in any gaps that are able to be filled. By publication time of this paper, we believe that any missing data that can be filled will have been done so for datasets provided directly by Baumgartner up to February 2016. Data shared from any other source may still include these missing observations. Since any missing data that Reddit does not provide will still be missing from the corrected datasets, we encourage researchers to check the integrity of your data when publishing results from this dataset.

More widely, the case of this so-called complete dataset draws attention to the risks to validity from research cultures that move fast to produce new results when new data is released. While many researchers have utilized Baumgartner's generous work on this Reddit dataset to investigate important questions, too few of us questioned a ``completeness'' statement that shouldn't have been accepted as truth. This dataset has numerous omissions, and those issues affect different research agendas with varying levels of severity. 

As researchers, we need to protect ourselves from the dazzling scale of large datasets. We encourage more people in Baumgartner's position to collect data, share it in an ethical manner, and contribute to knowledge through the research that it enables. It will not always be possible or reasonable to place strict methodological expectations upon such citizen scientists -- that responsibility lies firmly on academics. We hope this paper will encourage other researchers to test their assumptions and document data quality when conducting social scientific research with large datasets that they did not collect.

\subsection{Acknowledgements}

The authors wish to thank Jason Baumgartner for providing the dataset and for corresponding with us to check our analyses. We are also grateful to Haji Mohammad Saleem and Derek Ruths, for valuable discussion on the risks to machine learning research. Finally, we wish to thank Jack Hessel, Lillian Lee, David Mimno, and Chenhao Tan for providing a careful examination of their previous work, which they discuss in Appendix A.

\bibliography{biblio}
\bibliographystyle{aaai}

\appendix
\section{Appendix A}



\section*{Supplementary material}

saying "all posts"; similar problematic shorthand in follow-up work should not have been used.
4. Even after Jason's latest re-scraping, though, there still are some inexplicable findings that call into question the possibility of absolute completeness of a corrected version of this dataset. Here are few observations we have made based on a reading of your work and some previous analyses we had conducted.
    a. In the new reddit 0-10M post scrape, only 955K (9.5%) of posts are found. Here are some (base36, base10) submissions that were found within the first 10M ids: (2pjh7, 4550875), (2pjhg, 4550884), (2pjhq, 4550894); under the one-id-per-post metric, there clearly are still a significant number of missing posts -- perhaps up to 90%. If one attempts to access posts in between these values (e.g., (2pjh8, 455086)) either via the [browser][3] or via the API, a 404 error is returned (as of 3/9/2018). We've run into similar questions before, and attempted to find out why posts might be missing.
    b. To be concrete: after reading your work, we conducted a test with our own set. How many of the "gaps" in our data could be filled by re-querying the API? Using a random test set of 400 uniformly sampled post ids in the base36-sequential gaps in our set, we found that only 53/400 (13%) were accessible -- the remaining 347 (87%) posts were errors (we checked to see if this was reproducible in the browser -- for the posts we checked, it was). It's not entirely clear as to what this means. This observation invites many questions, e.g., are posts always assigned a continuous sequence of base36 ids? Can reddit moderators fully delete posts? Does the availability of posts according to the API change over time? Were the 13% of found posts unavailable for some reason in 2014? Have practices remained consistent since 2007? Do these potentially missing posts mean that future reddit studies are not to be trusted?
    c. Among the 347 failed API queries, there was a mixture of 404 and 403 errors (specifically: 289 404 errors (generally -- these were from earlier posts), 58 403 errors (generally -- these were from later posts)). According to the reddit python API documentation, a 403 error can be caused by a variety of circumstances, e.g., if the subreddit is private and the requesting user doesn't have access to it. According to the reddit python API, a 404 error occurs if the resource does not exist; it's plausible that a non-existent resource could be caused by an id that was never allocated, though it's hard to say why this happens. We have attempted to search through cached internet resources in search of missing posts (e.g., the internet archive) to no avail.
    d. We have also noticed some odd patterns in early reddit ids that could plausibly be indicative of future discontinuities, particularly if the id assignment scheme has remained consistent over time. For example, when sorting the very first posts made to reddit according to Jason's latest scrape by the created_utc field, the post ids up to post 28128 appear to be in base 10 (in base36, they are highly

---

[3] http://reddit.com/r/reddit.com/comments/1l3tm/broken_link/

inconsistent). Then, the next post's id is "sqh" which would be post 37241, made 71 minutes after. Scrolling through this list reveals many more discontinuities; in time, the following post sequence appears: 101x, 5yb9e (200 seconds later), 5yb98 (300 seconds later), 102c (50 seconds later). These oddities continue, too; for instance, here is a [post](http://reddit.com/r/reddit.com/comments/1l3tm/broken_link/)[4] from the new re-scraping that is impossible -- the comment was made 4 years prior to the post.
   e. Anecdotally, it seems that most of the oddities in the dataset occur before 2008. Jan. 2008 is when Reddit started to allow users to create their own communities. We have found that most of the missing data after 2008 are due to 403 errors, which means that these submissions are likely not publicly available. As a result, we think that the Reddit dataset curated by Jason B. is amenable to studies focused on user-created communities, particularly in comparison to other publicly available datasets.

In closing, we think that your work has raised an interesting research question: given that it's generally impossible to know if one has collected a "complete" set of information (be it posts by a reddit user, a user's tweets, wikipedia talk comments, etc.) are there methods of evaluation that simultaneously allow us to ask interesting questions while making sure observations are insensitive to often-overlooked (and, in the case of reddit, potentially un-avoidable) scraping gaps? We would be happy to discuss this topic going forward!

And, one bureaucratic point: if you found this email (or followup conversations) helpful, we would be happy about that, but ask that if you were to acknowledge us in any paper or the like, please include a disclaimer that we do not necessarily agree with every point made in your paper. (Obviously, you don't have to acknowledge us at all if this wasn't helpful!)

Signed (alphabetically),

Jack Hessel
Lillian Lee
David Mimno
Chenhao Tan

---

[4] http://reddit.com/r/reddit.com/comments/1l3tm/broken_link/


\clearpage
\end{document}